\documentclass[twocolumn,aps,showpacs,prx]{revtex4-1}

\usepackage{amsmath,amssymb,graphics,epsfig,epstopdf,color,multirow,array,verbatim,ulem,braket,tabularx}
\usepackage[colorlinks,linkcolor=blue,citecolor=blue,urlcolor=blue]{hyperref}
\usepackage{color}
\usepackage{graphicx}
\usepackage{epstopdf}
\usepackage{amsmath}
\usepackage{multirow}
\usepackage{ulem}
\usepackage{orcidlink}

\begin{document}

\title{Exploring charge and spin fluctuations in infinite-layer cuprate SrCuO$_{2}$ from a phonon perspective}

\author{Xin Du\orcidlink{0000-0003-1918-7568}$^1$}
\author{Pei-Han Sun\orcidlink{0000-0003-0177-8124}$^1$}
\author{Ben-Chao Gong\orcidlink{0000-0003-2203-6998}$^1$}
\author{Jian-Feng Zhang\orcidlink{0000-0001-7922-0839}$^2$}
\author{Zhong-Yi Lu\orcidlink{0000-0001-8866-3180}$^{1,3}$}\email{zlu@ruc.edu.cn}
\author{Kai Liu\orcidlink{0000-0001-6216-333X}$^{1,3}$}\email{kliu@ruc.edu.cn}

\affiliation{$^1$Department of Physics and Beijing Key Laboratory of Opto-electronic Functional Materials $\&$ Micro-nano Devices, Renmin University of China, Beijing 100872, China \\
$^2$Institute of Physics, Chinese Academy of Sciences, Beijing 100190, China \\
$^3$Key Laboratory of Quantum State Construction and Manipulation (Ministry of Education), Renmin University of China, Beijing 100872, China}

\date{\today}

\begin{abstract}

The infinite-layer cuprate $A$CuO$_2$ ($A=$ Ca, Sr, Ba) has the simplest crystal structure among numerous cuprate superconductors and can serve as a prototypical system to explore the unconventional superconductivity. Based on the first-principles electronic structure calculations, we have studied the electronic and magnetic properties of the infinite-layer cuprate SrCuO$_{2}$ from a phonon perspective. We find that interesting fluctuations of charges, electrical dipoles, and local magnetic moments can be induced by the zero-point vibrations of phonon modes in SrCuO$_{2}$ upon the hole doping. Among all optical phonon modes of SrCuO$_{2}$ in the antiferromagnetic N\'{e}el state, only the $A_{1}$$_g$ mode that involves the full-breathing O vibrations along the Cu-O bonds can cause significant fluctuations of local magnetic moments on O atoms and dramatic charge redistributions between Cu and O atoms. Notably, due to the zero-point vibration of the $A_{1g}$ mode, both the charge fluctuations on Cu and the electrical dipoles on O show a dome-like evolution with increasing hole doping, quite similar to the experimentally observed behavior of the superconducting $T_c$; in comparison, the fluctuations of local magnetic moments on O display a monotonic enhancement along with the hole doping. Further analyses indicate that around the optimal doping, there exist a large softening in the frequency of the $A_{1g}$ phonon mode and a van Hove singularity in the electronic structure close to the Fermi level, suggesting potential electron-phonon coupling. Our work reveals the important role of the full-breathing O phonon mode playing in the infinite-layer SrCuO$_{2}$, which may provide new insights in understanding the cuprate superconductivity.

\end{abstract}


\maketitle

\section{INTRODUCTION}

Since the discovery of high-temperature superconductivity in La$_{2}$$_{-}$$_x$Ba$_x$CuO$_{4}$ \cite{1}, extensive studies have been conducted to search for new materials with high superconducting transition temperatures ($T_c$'s) and to explore their underlying superconducting mechanism \cite{2, 3, 4, 5, 6}. However, unlike in conventional superconductors, the Bardeen-Cooper-Schrieffer (BCS) theory \cite{7} concerning the electron-phonon coupling fails to explain the high-$T_c$ superconductivity in cuprates. Although many experimental and theoretical works have investigated the roles of spin, charge, orbital, and nematicity in the strongly correlated cuprate superconductors \cite{8, 9, 10, 11}, no consensus has yet been reached on their superconducting mechanism. Up to now, it is widely accepted that the spin-mediated $d$-wave pairing symmetry is in a dominant position in cuprate superconductors \cite{12}. On the other hand, a host of experiments have indicated that phonon is also an essential factor in understanding the superconductivity of cuprates \cite{13, 14, 15, 16, 17, 18, 19}. In the recent studies on the Ba$_{2}$$_{-}$$_x$Sr$_x$CuO$_{3}$$_{+}$$_\delta$ superconductor, an additional strong near-neighbor attraction is attributed to the electron-phonon ($el$-$ph$) interaction \cite{20,21}.

The crystal structures of cuprate superconductors usually consist of alternating two-dimensional (2D) CuO$_{2}$ planes and buffer layers \cite{22}. Among various cuprates, the infinite-layer compounds $A$CuO$_{2}$ ($A$ = alkali earth ions) have the simplest structure, which is beneficial for exploring the superconducting mechanism. Early experiments obtained the infinite-layer Ca$_{1-x}$Sr$_x$CuO$_2$ ($x \approx$ 0.1) at ambient pressure \cite{Roth88} and $A$CuO$_{2}$ ($A$ = Ca$_{2/3}$Sr$_{1/3}$-Sr-Ba$_{1/3}$Sr$_{2/3}$) at 6 GPa \cite{24}. Subsequent study reported that the infinite-layer cuprate (Ca$_{0.3}$Sr$_{0.7}$)$_{0.9}$CuO$_{2}$ synthesized under pressure possesses a superconducting $T_c$ as high as 110 K \cite{23}. Later on, $A$CuO$_2$ epitaxial films were also successfully prepared on the perovskite substrates \cite{Shen12, Xue18, Xue20}. Similar to other cuprate superconductors, the properties of infinite-layer cuprate compounds can be effectively modulated by the pressure and the charge doping \cite{25, 26, 27, song22}. While the pressure has an immediate regulation on the lattice (phonon), the doping directly affects the charge and spin characteristics. Here, we focus on the hole-doped SrCuO$_{2}$, whose superconducting $T_c$ increases monotonically from 91 K at ambient pressure to $\sim$110 K at 8 GPa \cite{25}. In consideration of our previous study that reveals the prominent spin-phonon coupling in FeSe \cite{28}, there is a chance to inspect the phonon effects on the charge and spin degrees of freedom in the infinite-layer SrCuO$_{2}$, which may provide helpful clues on its superconducting mechanism. 

In this work, we have performed the systematic first-principles calculations to investigate the lattice dynamics, electronic structure, and magnetic properties of the infinite-layer SrCuO$_{2}$ under different hole doping levels. We uncover a nonnegligible effect of the $A_{1}$$_g$ phonon mode, i.e. the full-breathing O vibrations along the Cu-O bonds, on modulating the charge and spin fluctuations in SrCuO$_{2}$. With increasing hole doping, the charge fluctuations on Cu atoms and the electrical dipole variations on O atoms induced by the zero-point vibrations of the $A_{1}$$_g$ phonon both show a dome-like evolution, quite similar to the behavior of the superconducting $T_c$. Meanwhile, there exist remarkable local magnetic moment fluctuations on O atoms and unusual softening of the $A_{1}$$_g$ phonon mode around the optimal hole doping. Our computational results on the phonon effects in SrCuO$_{2}$ call for future experimental examination.

\section{COMPUTATIONAL DETAILS}

The structural, electronic, and magnetic properties of the pristine and hole-doped SrCuO$_{2}$ were studied with the density functional theory (DFT) calculations \cite{29,30} as implemented in the Vienna \textit{ab initio} simulation package (VASP) \cite{31}. The projector augmented wave (PAW) \cite{32} potentials with the valence electrons of 4$s^2$4$p^6$5$s^2$, 3$d^{10}$4$s^1$, and 2$s^2$2$p^4$ were used for the Sr, Cu, and O atoms, respectively. The generalized gradient approximation (GGA) of Perdew-Burke-Ernzerhof (PBE) formalism \cite{33} was adopted for the exchange-correlation functionals. The kinetic energy cutoff of the plane wave basis was set to 550 eV. A 12$\times$12$\times$20 Monkhorst-Pack {\bf k}-mesh \cite{34} was used for the $\sqrt{2}$$\times$$\sqrt{2}$$\times$1 supercell. To describe the strong correlation effect among Cu 3$d$ electrons, a Hubbard interaction $U$ was included in the calculations. Based on the linear response theory calculations \cite{35}, the effective Hubbard $U$ was determined to be 8 eV, with which both the antiferromagnetic Mott insulator feature and the local magnetic moments on Cu (Figs. S1 and S4 in the Supporting Information (SI) \cite{36}) were well reproduced for SrCuO$_{2}$ and CaCuO$_{2}$ \cite{37, 38, 39}. The hole doping effect was simulated by changing the total number of electrons in the supercell with a compensating Jellium background. The phonon spectra were obtained via the frozen phonon approach \cite{40}. To simulate the zero-point vibration of a specific phonon mode $s$, the atoms were displaced away from their equilibrium positions along two opposite directions in the normal-mode coordinate \cite{28}, both having the potential energy of $\hbar\omega_s$/2.

\section{Results and discussion}

\begin{figure}[!t]
\includegraphics[angle=0,scale=0.25]{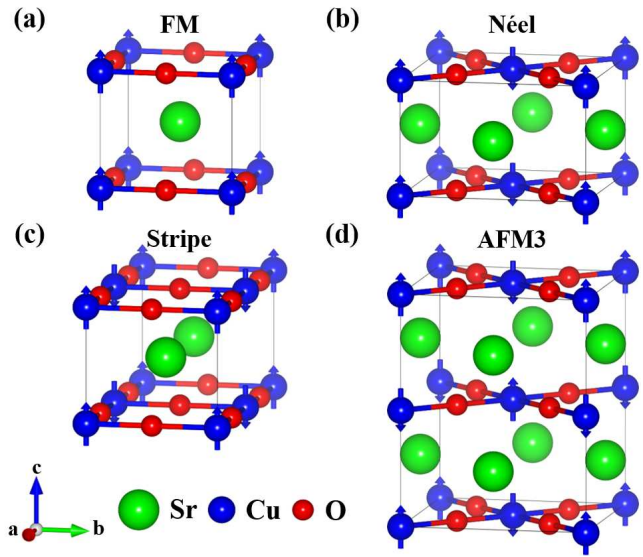}
\caption{(Color online) Crystal structures and typical spin configurations of SrCuO$_{2}$: (a) ferromagnetic (FM), (b) antiferromagnetic (AFM) N\'{e}el, (c) stripe AFM, and (d) AFM3 states. The green, blue, and red balls represent the Sr, Cu, and O atoms, respectively. The blue arrows on Cu atoms denote different spin polarizations.}
\label{fig1}
\end{figure}

\subsection{Crystal structure and magnetic configuration}

The crystal structure of infinite-layer cuprate SrCuO$_{2}$ is shown in Fig. \ref{fig1}. In each CuO$_{2}$ plane a Cu atom is connected with four O atoms forming a 2D square lattice, while the CuO$_{2}$ and Sr planes stack alternatively along the $c$ direction. To determine the magnetic ground state of SrCuO$_{2}$, we considered the nonmagnetic (NM) state, the ferromagnetic (FM) state, and three typical antiferromagnetic (AFM) states (N\'{e}el, stripe, and AFM3). The NM and FM states were calculated in the primitive cell, while the AFM N\'{e}el, stripe AFM, and AFM3 states were simulated with the $\sqrt{2}$$\times$$\sqrt{2}$$\times$1, 2$\times$1$\times$1, and $\sqrt{2}$$\times$$\sqrt{2}$$\times$2 supercells, respectively (see Fig. \ref{fig1}). In the AFM N\'{e}el state, the intralayer nearest-neighbor coupling between Cu spins is AFM and the interlayer coupling is FM (Fig. \ref{fig1}b). In the stripe AFM state, Cu spins in each CuO$_{2}$ plane are antiferromagnetically coupled along $a$ axis and ferromagnetically coupled along $b$ axis (Fig. \ref{fig1}c). As for the AFM3 state, both intralayer and interlayer nearest-neighbor couplings are antiferromagnetic (Fig. \ref{fig1}d).

\begin{table}[!b]
\caption{Calculated lattice constants, total energies (relative to that of the AFM N\'{e}el state), and local magnetic moments on Cu atoms for the typical magnetic states of SrCuO$_{2}$.}
\begin{center}
\begin{tabular*}{8cm}{@{\extracolsep{\fill}} ccccccc}
\hline \hline
States & Lattice constant & Energy & $M_{\rm Cu}$ \\
 &({\AA}) & (meV/f.u.) & ($\mu$${\rm _B}$) \\
\hline
NM & $a$ = $b$ = 3.94, $c$ = 3.48 & +279.4 & / \\
FM & $a$ = $b$ = 3.96, $c$ = 3.47 & +153.5 & 0.76  \\
N\'{e}el & $a$ = $b$ = 3.95, $c$ = 3.48 & 0.0 & 0.61  \\
Stripe & $a$ = $b$ = 3.97, $c$ = 3.47 & +64.6 & 0.68  \\
AFM3 & $a$ = $b$ = 3.95, $c$ = 3.48 & -1.1 & 0.61  \\
\hline
\hline
\label{table1}
\end{tabular*}
\end{center}
\end{table}

\begin{figure*}[!t]
\includegraphics[angle=0,scale=0.32]{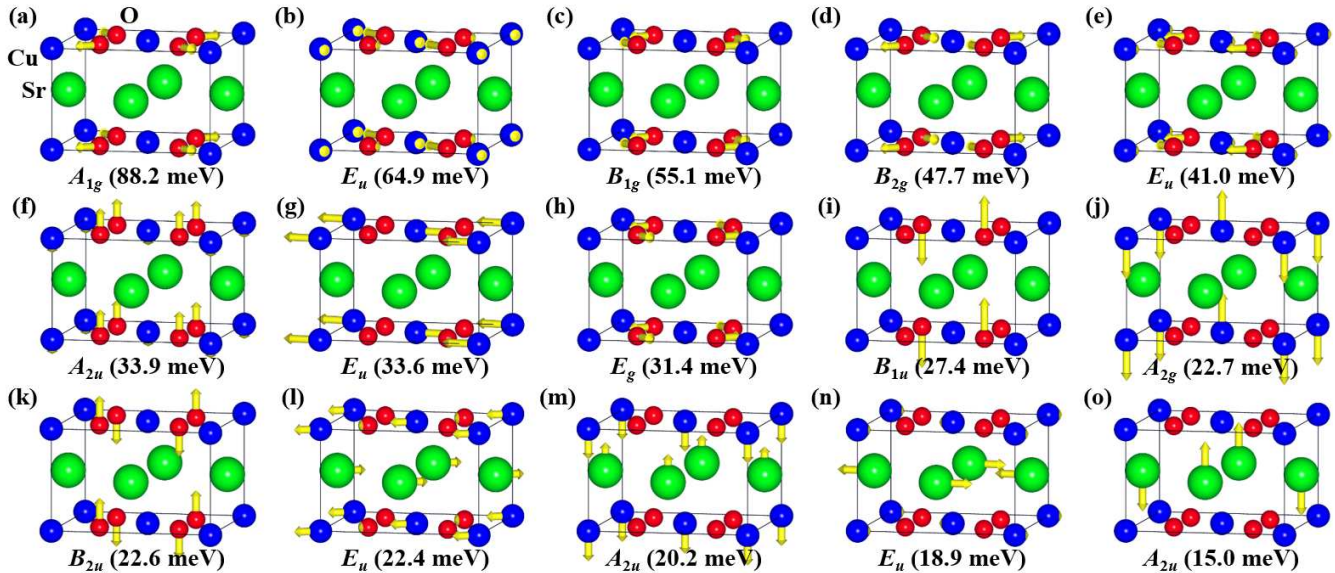}
\caption{(Color online) Atomic displacement patterns for fifteen optical phonon modes of the hole-doped SrCuO$_{2}$ in the AFM N\'{e}el state. The yellow arrows denote the directions and amplitudes of the atomic vibrations. The corresponding symmetries and frequencies of the phonon modes are also labeled in each panel, while the $E_u$ and $E_g$ modes are both doubly degenerate. }
\label{fig2}
\end{figure*}

Table \ref{table1} shows the calculation results for these typical magnetic states of SrCuO$_{2}$ with Hubbard $U_{\rm eff}$ = 8 eV. The lattice constants are referred to those of the primitive cells, which agree very well with the experimental values ($a$ = $b$ = 3.926 \AA, $c$ = 3.432 \AA) \cite{24, 27}. As expected, the AFM N\'{e}el states of SrCuO$_{2}$ is more stable than other magnetic states, while its energy is 279.4 meV, 153.5 meV, and 64.6 meV per formula unit (f.u.) lower than those of the NM, FM, and stripe AFM states, respectively. It is notable that both lattice constants and total energies of the AFM N\'{e}el state and the AFM3 state are quite similar, which originates from the weak interlayer coupling between the CuO$_{2}$ planes \cite{41, 42}. For a convenient comparison with other studies \cite{39, 42}, we chose the AFM N\'{e}el state as the magnetic ground state in the following calculations. In this AFM N\'{e}el state, the local magnetic moment on Cu is 0.61 $\mu_\text{B}$. Moreover, the adoption of different $U$ values (5, 6.258, 7.5, or 8 eV) according to previous references \cite{39, 43, 44} does not change the relative stabilities of these magnetic states (Table S1 of SI \cite{36}).

\subsection{Phonons and their influence on the electronic and magnetic properties of the pristine SrCuO$_{2}$}

Our previous study on FeSe has suggested that the spin-phonon coupling may have an important impact on the iron-based superconductors \cite{28}, here we further explore the phonon effect on the electronic and magnetic properties of the infinite-layer cuprate SrCuO$_{2}$.

Firstly, we studied the nonmagnetic state of SrCuO$_{2}$ in the primitive cell, whose atomic displacement patterns of the phonon modes with the corresponding frequencies and symmetries are displayed in Fig. S2 of SI \cite{36}. There are five infrared active modes (3$E_u$ and 2$A_{2}$$_u$) and no Raman active mode, and the frequency range of these optical phonon modes is 143-459 cm$^{-1}$, which are in good accord with previous experiments \cite{45} and also validate the accuracy of our calculations.

Then we explored the AFM N\'{e}el ground state of undoped SrCuO$_{2}$ in a $\sqrt{2}$$\times$$\sqrt{2}$$\times$1 supercell. There are fifteen nonequivalent optical phonon modes, whose atomic displacement patterns are shown in Fig. S3 of SI \cite{36} in a sequence with decreasing frequencies. Our band structure calculations show that it is an AFM insulator with a bandgap of 0.8 eV (Fig. S4a-b \cite{36}). To further study the effect of these phonon modes on the electronic and magnetic properties, we displaced the atoms in each specific phonon mode $s$ with the corresponding zero-point energy of $\hbar\omega$$_s$/2. Because the atoms have two opposite directions in the normal-mode coordinates (defined as '+' and '$-$'), there are two displacement patterns for each phonon mode. By comparing the differences between these two cases, we obtained the variations in charges ($\vert$$\Delta$$\rho$$\vert$ = $\vert$$\rho$$_+$ $-$ $\rho$$_-$$\vert$) and local magnetic moments ($\vert$$\Delta$$M$$\vert$ = $\vert$$M_+$ $-$ $M_-$$\vert$) on Cu and O atoms for all fifteen phonon modes (Fig. S4c-d \cite{36}). Unexpectedly, the local moment variations on all Cu and O atoms due to the positive ('+') and negative ('$-$') displacements of the phonon modes do not exceed 0.03 $\mu$${\rm _B}$, suggesting the weak phonon effect on the local moments in the pristine (undoped) SrCuO$_{2}$. Although small in magnitude (Fig. S4d \cite{36}), the most obvious variations in local moments are mainly induced by the high-frequency in-plane O vibrations (Fig. S3a-e \cite{36}).

\subsection{Phonon effects on the electronic and magnetic properties of the hole-doped SrCuO$_{2}$}

\begin{figure}[!t]
\includegraphics[angle=0,scale=0.27]{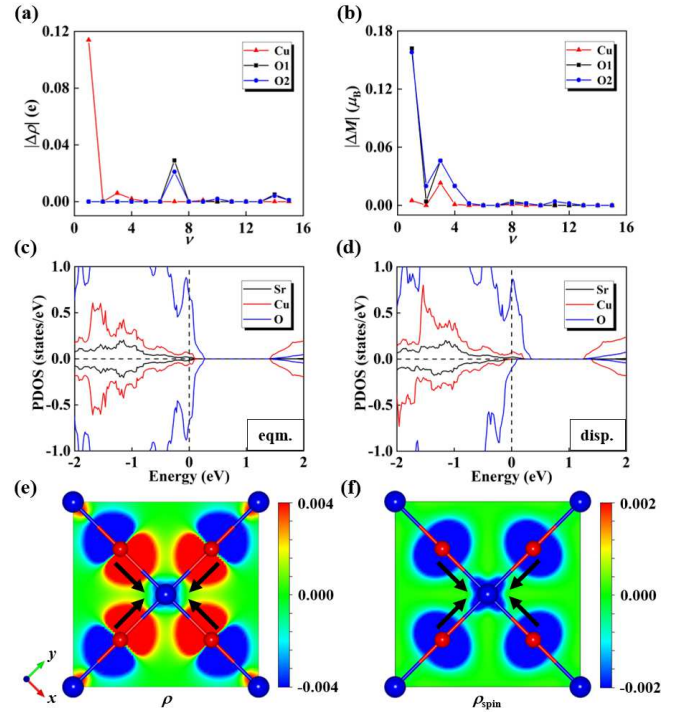}
\caption{(Color online) Variations of (a) charges and (b) local magnetic moments on the Cu and O atoms in the hole-doped (0.2 holes/Cu) SrCuO$_{2}$ induced by the zero-point vibrations of different optical phonon modes. The horizontal axis $\nu$ labels the No. of phonon mode in Fig. \ref{fig2}. Partial density of states (PDOS) of the hole-doped SrCuO$_{2}$ in the AFM N\'{e}el state for (c) the equilibrium structure and (d) the one with the zero-point vibrational displacement of the $A_{1}$$_g$ mode. (e) Differential charge density map and (f) differential spin density map for the hole-doped SrCuO$_{2}$ between the distorted structure and the equilibrium structure plotted on the (001) plane. The blue and red balls represent the Cu and O atoms, respectively. The $x$ and $y$ axes along the Cu-O bonds are defined to better describe the atomic orbitals. The black arrows denote the vibrational directions of O atoms. The color bars in (e) and (f) are in units of e/{\AA}$^3$.}
\label{fig3}
\end{figure}

Considering the important role of charge doping in the emergence of superconductivity in cuprates, for example the infinite-layer cuprate (Ca$_{0.3}$Sr$_{0.7}$)$_{0.9}$CuO$_{2}$ possessing a $T_c$ of 110 K \cite{23}, we next focused on the hole-doped SrCuO$_{2}$. We introduced 0.4 holes in the $\sqrt{2}$$\times$$\sqrt{2}$$\times$1 supercell for the AFM N\'{e}el state of SrCuO$_{2}$, which corresponds to a doping level of 0.2 holes/Cu. The atomic displacement patterns with corresponding symmetries for the phonon modes in the hole-doped SrCuO$_{2}$ are shown in Fig. \ref{fig2}, which are similar to those of the pristine SrCuO$_{2}$ (Fig. S3 in SI \cite{36}). Among the fifteen nonequivalent phonon modes, there are four Raman active, eight infrared active, and three silent modes. Those four Raman active modes are the $A_{1}$$_g$, $B_{1}$$_g$, $B_{2}$$_g$, and $E_g$ modes, all of which involve the in-plane vibrations of O atoms (Figs. \ref{fig2}a, \ref{fig2}c, \ref{fig2}d, and \ref{fig2}h), while the three silent modes ($B_{1}$$_u$, $A_{2}$$_g$ and $B_{2}$$_u$) show out-of-plane vibrations of O and Cu atoms (Figs. \ref{fig2}i, \ref{fig2}j and \ref{fig2}k). Further calculations indicate that only the $A_{1}$$_g$ mode, which relates to the full-breathing O in-plane vibrations along the Cu-O bonds (Fig. \ref{fig2}a), causes significant changes in the charges on Cu atoms (Fig. \ref{fig3}a) and the local magnetic moments on O atoms (Fig. \ref{fig3}b), respectively. Here SrCuO$_{2}$ is different from FeSe, in the latter the local moments on Fe being very sensitive to the vertical Se vibrations perpendicular to the Fe-Fe plane \cite{28}.

To intuitively display the variation of local magnetic moments induced by the zero-point vibrations of all phonon modes in the hole-doped SrCuO$_{2}$, we also plotted the real-space spin density maps in the (001) and (110) planes (Figs. S5 and S6 \cite{36}). Consistent with the above results in Fig. \ref{fig3}b, it is obvious that only the full-breathing $A_{1}$$_g$ mode causes prominent spin density fluctuations around the O atoms (Fig. \ref{fig3}f).

In order to investigate the effect of the full-breathing $A_{1}$$_g$ mode on the electronic properties of the hole-doped SrCuO$_{2}$, we plotted the band structure (Fig. S7 \cite{36}) as well as the partial density of states (PDOS) in the equilibrium structure and the one with atomic displacements (Fig. \ref{fig3}c-d). The spin-up and spin-down parts of the DOS in the hole-doped SrCuO$_{2}$ become asymmetric due to the O displacements in the $A_{1}$$_g$ phonon mode. By comparing the PDOSs of different atomic orbitals in the equilibrium and displaced structures, we find that the occupation changes are contributed by the O $p_{x/y}$ and Cu $d_{x^2-y^2}$ orbitals (Fig. S8 \cite{36}). 

Another notable feature in PDOS is that the changes of O orbitals are much greater than those of Cu around the Fermi level (Figs. \ref{fig3}c-d), suggesting that the charge fluctuations may mainly come from the O atoms, which seems to be controversial with the results in Fig. \ref{fig3}a. To better understand this point, we further plotted the differential charge densities between the structure with atomic displacements and the one at equilibrium position (Fig. \ref{fig3}e), which shows that the real-space charge variations are indeed the most dramatic around the O atoms. Nevertheless, due to the antisymmetry between the positive and negative variations on charges around the O atom, the integrated charge fluctuation on each O atom approaches zero, leaving that on the Cu atom looking prominent (Fig. \ref{fig3}a). It is worth noting that the antisymmetric charge variations around each O atom also cause the centers of the positive and negative charges to not coincide, indicating that there exist dynamical electrical dipoles induced by the atomic displacements of the $A_{1}$$_g$ phonon mode (Fig. \ref{fig3}e).

\subsection{Electronic structure, magnetic properties, and phonon spectra as functions of hole doping in SrCuO$_{2}$}

The cuprate superconductors usually show a dome-like evolution of the superconducting $T_c$ with the hole doping. We then studied the electronic structure, magnetic properties, and phonon spectra of SrCuO$_{2}$ as functions of the hole doping concentration.

The total energy calculations of the NM, FM, AFM N\'{e}el, and stripe AFM states of SrCuO$_{2}$ were carried out under the doping levels ranging from 0.05 holes/Cu to 0.25 holes/Cu. As shown in Table S2 \cite{36}, the AFM N\'{e}el state is the most stable one, but the energy differences among the above magnetic states decrease with the hole doping. To describe the magnetic interactions in SrCuO$_{2}$, we calculated the nearest-neighbor exchange $J_1$ and next-nearest-neighbor exchange $J_2$ between Cu spins based on an effective Heisenberg model \cite{46, 47, 48}. For pristine SrCuO$_{2}$, the calculated $J_1$ = 153 meV and $J_2$ = 12 meV are quite similar to the reported values of the isostructural CaCuO$_{2}$ ($J_1$ = 155 or 146 meV and $J_2$ = 10.3 meV) \cite{49, 50, 51}. With the increase of hole doping, both the decreasing $J_1$ and the sign change of $J_2$ indicate the tendency of magnetic frustration (Table S3 \cite{36}).

\begin{figure}[!t]
\includegraphics[angle=0,scale=0.28]{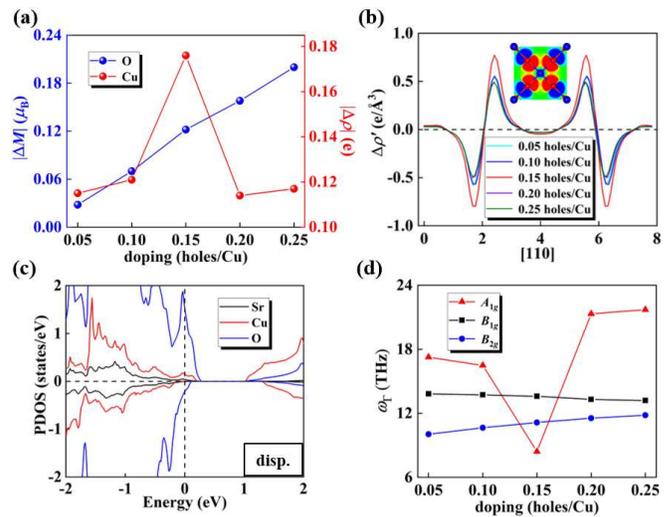}
\caption{(Color online) (a) Charge fluctuations on the Cu atoms and local magnetic moment fluctuations on the O atoms as functions of hole doping. (b) Differential charge densities along the [110] direction induced by the atomic displacements due to the zero-point vibrations of $A_{1}$$_g$ phonon mode as a function of hole doping. The inset shows the differential charge density map at the doping of 0.15 holes/Cu. (c) PDOS of SrCuO$_{2}$ under the atomic displacements of the zero-point vibration of the $A_{1}$$_g$ mode at the doping of 0.15 holes/Cu. (d) Phonon frequencies at the Brillouin zone center for the $A_{1}$$_g$, $B_{1}$$_g$, and $B_{2}$$_g$ phonon modes of SrCuO$_{2}$ in the AFM N\'{e}el state as a function of hole doping.}
\label{fig4}
\end{figure}

The phonon effects induced by the zero-point vibrations on the electronic and magnetic properties of SrCuO$_{2}$ under different hole doping levels are plotted in Figs. S9-S11 of SI \cite{36}. The results indicate that it is still the full-breathing $A_{1}$$_g$ mode that causes the most obvious charge fluctuations on the Cu atoms and spin fluctuations on the O atoms, respectively. For better visualization, Fig. \ref{fig4}a plots these variations on the Cu and O atoms due to the zero-point vibration of the $A_{1}$$_g$ phonon mode as functions of hole doping. With the increasing hole doping, the charge fluctuations around the Cu atoms first increase and then decrease in a dome-like manner, while the highest value appears near the doping of 0.15 holes/Cu. In contrast, the local magnetic moment fluctuations on the O atoms increase monotonically with the hole doping. To analyze the real-space charge fluctuations in detail, we plotted the differential charge density induced by the atomic displacements of the $A_{1}$$_g$ mode along the [110] direction in the doping range of 0.05-0.25 holes/Cu (Fig. \ref{fig4}b). Around each O atom, the centers of the positive and negative charges do not coincide, resulting in the dynamical electrical dipoles (inset of Fig. \ref{fig4}b and Fig. \ref{fig3}e). It can be seen that the evolution of the electrical dipoles around the O atoms with the hole doping is consistent with that of the charge fluctuations around the Cu atoms (Fig. \ref{fig4}a), which has a maximum value near the doping of 0.15 holes/Cu (Fig. \ref{fig4}b and Fig. S11 of SI \cite{36}).

From the electronic band structures of SrCuO$_{2}$ under different doping concentrations (Fig. S12 in SI \cite{36}), it can be seen that a hole doping around 0.15 holes/Cu can induce the change of Fermi surface topology, namely the Lifshitz transition. This critical doping is somewhat different from the proposed value (0.24 holes/Cu) in CaCuO$_{2}$ \cite{26}, but is similar to the well-accepted optimal doping ($\sim$0.16 holes/Cu) for other high-$T_c$ cuprates \cite{2, 52, 53}. Moreover, since the O displacements due to the zero-point vibration of the $A_{1}$$_g$ mode can also shift the relative positions of atomic orbitals with respect to the Fermi level (Figs. \ref{fig3}c and \ref{fig3}d), we plotted in Fig. \ref{fig4}c the PDOS of SrCuO$_{2}$ under the O displacements of the $A_{1}$$_g$ mode at the doping of 0.15 holes/Cu. Clearly, both the van Hove singularities in the PDOSs of the O and Cu atoms are very close to the Fermi level at this doping concentration (Fig. \ref{fig4}c and Figs. S13,14 in the SI \cite{36}). These results indicate that with the aid of hole doping at an optimal level, the $A_{1}$$_g$ phonon mode in SrCuO$_{2}$ can manifest its largest effect on the electronic structures.

To further explore the evolution of phonon modes with the hole doping, we focused on three high-frequency phonon modes ($A_{1}$$_g$, $B_{1}$$_g$, and $B_{2}$$_g$) related to the in-plane symmetric O vibrations around the Cu atoms (Fig. \ref{fig2}). The calculated phonon frequencies at the Brillouin zone center (Fig. \ref{fig4}d) show that only the frequency of the $A_{1}$$_g$ mode displays a nonmonotonic behavior in an anti-dome shape with the increasing hole doping, whereas the frequencies of the other two modes show linear evolutions. Remarkably, at the doping of 0.15 holes/Cu, the $A_{1}$$_g$ mode shows dramatical softening and has the minimum frequency. Given that at this doping concentration the van Hove singularities in PDOS is also close to the Fermi level (Fig. \ref{fig4}c), we suggest that there may exist strong electron-phonon coupling between the electrons in the CuO$_{2}$ plane and the $A_{1}$$_g$ phonon mode that involves the full-breathing O vibrations.

\section{DISCUSSION AND SUMMARY}

The electronic and magnetic properties of SrCuO$_{2}$ have been investigated from a phonon viewpoint. First, the zero-point vibrations of phonon modes can induce concurrent fluctuations of charges on the Cu as well as electrical dipoles and local magnetic moments on the O in the hole-doped SrCuO$_{2}$ around the optimal doping (Figs. \ref{fig3} and \ref{fig4}), which suggest a non-negligible role of phonon playing in the superconductivity. Second, among the phonon modes, the full-breathing $A_{1}$$_g$ mode causes the largest fluctuations of charges and local magnetic moments in the doped SrCuO$_{2}$ (Fig. \ref{fig3} and Figs. S9-S11 in the SI \cite{36}), revealing the importance of the symmetric O vibrations along the Cu-O bonds (Fig. \ref{fig2}a). Third, at the optimal doping ($\sim$ 0.15 holes/Cu) where the Lifshitz transition in the electronic structure of SrCuO$_{2}$ is ready to occur (Fig. S12 in the SI \cite{36}), the O displacements related to the $A_{1}$$_g$ mode can alter the relative position of the van Hove singularity in the PDOS with respect to the Fermi level (Fig. \ref{fig4}c), suggesting potential strong electron-phonon coupling. Last but not least, the comparison among the phonon frequencies of $A_{1}$$_g$, $B_{1}$$_g$, and $B_{2}$$_g$ modes show that only the $A_{1}$$_g$ mode exhibits apparent softening around the optimal doping (Fig. \ref{fig4}d), implying its close relation with the unconventional superconductivity in the hole-doped SrCuO$_{2}$.

Our computational results and theoretical analyses on SrCuO$_{2}$ have identified the interesting role that the full-breathing O vibrations in the CuO$_{2}$ plane plays in inducing charge/spin fluctuations and modulating the van Hove singularity positions. Note that previous studies on unconventional superconductors have also revealed the important effects of phonons. For example, a number of ARPES experiments and theoretical studies have confirmed that some special phonon modes (e.g. the $B_{1}$$_g$ phonon related to the out-of-plane O vibrations of CuO$_{6}$ octahedron) are associated with the $d$-wave pairing, which highlights the importance of $el$-$ph$ coupling in cuprate superconductors \cite{16, 54, 55}. Recently, the resonant inelastic x-ray scattering (RIXS) experiments on Bi$_{2}$Sr$_{1.4}$La$_{0.6}$CuO$_{6}$$_{+}$$_\delta$ observed charge density wave excitations accompanied by phonon anomalies \cite{18}. Interestingly, not only for cuprates, there are also evidences for the phonon effects in iron-based superconductors. For instance, the enhanced superconducting $T_c$ of FeSe monolayer grown on SrTiO$_{3}$ \cite{56} is proposed to be closely related with the interaction between the FeSe electrons and the SrTiO$_{3}$ phonons \cite{57}. In addition, our previous study on FeSe also suggested that the phonon mode involving the vertical vibrations of Se atoms is intimately connected to the spin fluctuations \cite{28}. All above theoretical and experimental studies suggest that phonons are indeed indispensable for a complete understanding the unconventional superconductivity.

To summarize, we have studied the variations of the electronic and magnetic properties induced by the zero-point vibrations of phonon modes in the pristine and hole-doped SrCuO$_{2}$ based on the spin-polarized density functional theory calculations. Referring to the dome-shape phase diagram of the superconducting $T_c$ with the hole doping, our calculations on SrCuO$_{2}$ reveal the similar doping dependences of the charge fluctuations on the Cu atoms and the electrical dipoles around the O atoms due to the full-breathing O vibrations of the $A_{1}$$_g$ phonon mode. We also find the enhanced local moment fluctuations on the O atoms, the emerging Lifshitz transition in the band structure, as well as the apparent softening of the $A_{1}$$_g$ phonon mode at the optimal hole doping. Our work reveals the important role of the full-breathing O phonon mode in the infinite-layer SrCuO$_{2}$, which provides an interesting microscopic picture toward unraveling the mystery of cuprate superconductivity.

\begin{acknowledgments}

This work was supported by the National Key R\&D Program of China (Grants No. 2022YFA1403103 and No. 2019YFA0308603), the Beijing Natural Science Foundation (Grant No. Z200005), and the National Natural Science Foundation of China (Grants No. 12174443 and No. 11934020). J.Z. was supported by the Project funded by China Postdoctoral Science Foundation (Grant No. 2022M723355). Computational resources have been provided by the Physical Laboratory of High Performance Computing at Renmin University of China and the Beijing Super Cloud Computing Center.

\end{acknowledgments}





\end{document}